\documentclass[useAMS,usenatbib]{mn2e}

\usepackage[psamsfonts]{amssymb}
\usepackage[dvips]{graphicx}
\usepackage{amsmath,alltt}
\usepackage{multirow}
\usepackage{rotating}
\usepackage{lscape}

\title[Small-N collisional dynamics II]{Small-N collisional dynamics II:  Roaming the realm of not-so-small-N}
\author[Leigh N. W. C., Geller A. M.]{Nathan W. C. Leigh$^{1,2}$, Aaron M. Geller$^{3,4}$
\thanks{E-mail: nleigh@ualberta.ca (NWCL), a-geller@northwestern.edu}\\
$^{1}$Department of Astrophysics, American Museum of Natural History, Central Park West and 79th Street, New York, NY 10024 \\
$^{2}$Department of Physics, University of Alberta, CCIS 4-183, Edmonton, AB T6G 2E1, Canada \\
$^{3}$Center for Interdisciplinary Exploration and Research in Astrophysics (CIERA) and Department of Physics and Astronomy, \\ Northwestern University,\
 2145 Sheridan Rd, Evanston, IL 60208, USA \\
$^{4}$Department of Astronomy and Astrophysics, University of Chicago, 5640 S. Ellis Avenue, Chicago, IL 60637}
\begin{document}

\pagerange{\pageref{firstpage}--\pageref{lastpage}} \pubyear{2011}

\maketitle

\label{firstpage}

\begin{abstract}
We develop a formalism for calculating probabilities for the outcomes 
of stellar dynamical interactions, based on results from $N$-body scattering experiments. 
We focus here on encounters involving up to six particles and calculate probabilities for direct 
stellar collisions; 
however our method is in principle valid for larger particle numbers.  
Our method relies on the binomial theorem, and is applicable to encounters involving any combination of 
particle radii.  We further demonstrate that our base model is valid to within a few percent 
for any combination of particle masses, provided the minimum mass ratio is within a factor of 
a few from unity.  
This method is particularly suitable for models of collisional systems involving large numbers of 
stars, such as globular clusters, old open clusters
and galactic nuclei, where small subsets of stars may regularly have very close encounters, 
and the direct integration of all such encounters is computationally expensive.  Variations of our 
method may also be used to treat other encounter outcomes, such as ejections and exchanges.
\end{abstract}

\begin{keywords}
gravitation -- binaries (including multiple): close -- globular clusters: general -- stars: kinematics and dynamics -- scattering -- methods: analytical.
\end{keywords}

\section{Introduction} \label{intro}


The three-body problem has a long history extending all the way back to \citet{newton1686}, 
yet it has never been fully solved analytically.  
Indeed, for the majority of the relevant parameter space and over sufficiently long timescales, the 
evolution is chaotic \citep{poincare1892}.  More recently, a number of different approaches have been invoked to 
study the problem \citep[e.g.][]{henon69,valtonen06}, 
and the most common approach today is to use computers to 
integrate the equations of motion directly (in some instances, using the secular approximation, e.g. \citealt{naoz14}).
And yet, in modern $N$-body simulations of star 
cluster evolution, it is the presence of binaries and especially higher-order multiples that poses 
the biggest computational challenge.  The time evolution of the orbits within multiple star systems, as 
well as the outcomes of chaotic close gravitational encounters involving multiples, 
requires very small time steps, relative to the crossing time of the entire cluster \citep[e.g.][]{hurley05,geller13}.  
This is particularly problematic, since simulations 
have shown that encounters involving binaries and triples can be crucial to not only the overall cluster 
evolution \citep[e.g.][]{hut83a,hut92}, but also the formation of exotic populations such as blue stragglers 
\citep{perets09,leigh11,geller13,naoz14} and even black holes \citep{mapelli14}.  What's more, observations 
have now revealed that higher-order multiple star systems are present in young star clusters in 
non-negligible numbers \citep[e.g.][]{leigh13}.

%

One avenue into alleviating the computational challenges of directly integrating these relatively low-$N$ 
systems (where $N$ is the number of particles involved in the interaction) is to assume that 
the final states specifying the outcomes of chaotic gravitational interactions are probabilistic, as opposed to deterministic, in so
far as how they relate to the initial encounter conditions.  For example, a strongly interacting three-body system 
typically breaks up to produce a binary and an escaping star.  \citet{monaghan76a} showed that, 
by making the assumption that the probability of a configuration is proportional to the 
associated volume in phase-space, the statistical properties of the remaining binary and the 
escaper can be predicted \citep{monaghan76b}.  This result rests heavily on Liouville's Theorem 
\citep{liouville38}, which states that, in a gravitionally-interacting system of particles, the 
total volume in phase-space is conserved in time.  

In this, and our previous paper \citep[Paper I;][]{leigh12}, we develop a probabilistic
interpretation of the outcomes of stellar encounters involving modest numbers of particles,
with an over-arching goal of connecting the small- and large-$N$ limits in bound configurations of finite-sized 
particles evolving under the influence of Newtonian gravity. 
In Paper I, we performed numerical scattering experiments involving single, binary and triple stars, using 
the \texttt{FEWBODY} code \citep{fregeau04}, and showed that the probability of a collision occuring during a chaotic 
gravitational interaction involving $N$ identical particles is proportional to $N^2$.  
Interestingly, this same $N$-dependence is what is predicted by the mean-free path approximation in the limit of very 
large $N$.  


In this paper, we delve further into the physical origin of this $N$-dependence, 
which relates to the binomial theorem, and explain how to exploit it to predict 
collision probabilities over a large range in the relevant parameter space of initial conditions with minimal 
computational expense.
Specifically, here we simulate
encounters involving \textit{different types of particles} 
(whereas in Paper I we limited our study to identical particles).  
We focus first on encounters involving particles with identical masses but different physical radii, and outcomes that result 
in a single direct collision between any two particles.  We then go on to consider encounters 
involving particles with different masses.  
In Section~\ref{method}, we present our method for calculating probabilities for the encounter 
outcomes (focusing on collisions, but with a discussion of how to expand these results to different outcomes).  
Our results are presented in Section~\ref{results}, which are then summarized and dicussed in Section~\ref{summary}.
 
\section{Method} \label{method}

In this section, we present our method for calculating collision probabilities for encounters 
involving different types of particles, as well as the numerical scattering experiments 
used to test our extrapolation technique.  We will focus first on encounters where the particle masses are all identical, 
but two different particle radii are possible.  However, we keep our approach as general as possible, 
since in principle it can be extended to any combination of particle numbers and types.
Later we relax the constraint of equal particle masses and begin a comparison to encounters involving particles 
of different masses. 

\subsection{Model} \label{model}

The key result from Paper I can be explained as follows.  First, we showed that the probability
of a collision occuring during a chaotic gravitational interaction involving $N$ identical particles is
proportional to $N^2$. (Note that a proportionality of $N(N-1)$ also fits the data; however we could not 
differentiate between these $N$-dependences at a statistically significant level.)  We briefly 
touched upon a connection between this result
and the large-$N$ limit via the mean free path approximation.  Specifically, in the
limit of very large $N$, the mean free path approximation predicts that the rate of collisions between any
two particles also scales as $N^2$.
The physical origin of this proportionality can be traced back to the binomial theorem.\footnote{More 
technically, the binomial theorem gives the different possible pairs of objects that could collide 
directly, and provides a good intuitive basis for building our model for the collision probability.  
Hence, our model implicitly assumes that other effects relevant to deciding the outcome of a chaotic 
gravitational encounter, and 
hence the collision probability, do not also depend on the number of interacting stars $N$.}

Briefly, the binomial theorem states that the number of subsets composed
of $k$ particles that can be constructed from a larger sample composed of $N$ particles is equal to:
\begin{equation}
\label{eqn:binthm}
{N \choose k} = \frac{N!}{(N-k)!k!}
\end{equation}
Thus, in our specific example of collisions, we want the probability of selecting any two particles
at random from our larger sample of $N$ \textit{identical} particles.  From the binomial theorem, the number of 
ways to select any two particles is:
\begin{equation}
\label{eqn:binthm2}
{N \choose 2} = \frac{N!}{(N-2)!2!} = \frac{N(N-1)}{2}
\end{equation}
This factor of $N(N-1)$ is proportional to the trend in collision probability as a function of $N$ particles that 
we found in Paper I (as stated above).  Furthermore, $N^2$ $\sim N(N-1)$ for $N \gg 1$, which is the 
scaling predicted by the mean free path approximation for collisions between any two particles.
That is, the probability of any two identical particles colliding 
is proportional to Equation~\ref{eqn:binthm2}.  In this way, the binomial theorem will form the backbone of any theory 
aiming to predict the relative 
probabilities of colliding two specific particles during a chaotic gravitational interaction.

What if the particles are not identical?  Specifically, let us consider the case of 
\textit{two} different particles, types 1 and 2.  Three different collision scenarios are now 
possible; a collision 
between two particles of type 1, a collision between two particles of type 2 and a 
collision between a particle of type 1 and a particle of type 2.  Thus, the collision 
probability can be broken up into the sum of three different terms, each corresponding to 
the probability of a specific collision event occurring:
\begin{equation}
\label{eqn:Pcoll1}
P_{\rm coll} = P_{\rm 11} + P_{\rm 12} + P_{\rm 22},
\end{equation}
where $P_{\rm 11}$ and $P_{\rm 22}$ are the probabilities for a collision between two 
particles of type 1 and two particles of type 2, respectively, and $P_{\rm 12}$ is the 
probability for a collision between a particle of type 1 and a particle of type 2.  Now, 
the dependence of each of these probability terms depends on the number of particles of 
each type $N_{\rm i}$, and this $N_{\rm i}$-dependence comes from the binomial theorem.  

Let $N_{\rm 1}$ and $N_{\rm 2}$ be the numbers of particles of type 1 and type 2, respectively, that are 
involved in the dynamical encounter.  Then, Equation~\ref{eqn:Pcoll1} can be re-written as:
\begin{equation}
\label{eqn:Pcoll2}
P_{\rm coll} = \alpha_{\rm 11}{N_{\rm 1} \choose 2} + \alpha_{\rm 12}{N_{\rm 1} \choose 1}{N_{\rm 2} \choose 1} + \alpha_{\rm 22}{N_{\rm 2} \choose 2},
\end{equation}
where $\alpha_{\rm 11}$, $\alpha_{\rm 12}$ and $\alpha_{\rm 22}$ are constants that depend on the 
total encounter energy $E$, the total angular momentum $L$ and the properties of the particles (mass 
and radius).  

We further assume that the particles have identical masses but two different radii are possible, namely 
$R_{\rm 1}$ and $R_{\rm 2}$ for particle types 1 and 2, respectively.  
We assume that the constant 
$\alpha_{\rm 12}$ can be found directly from the constants $\alpha_{\rm 11}$ and $\alpha_{\rm 22}$:
\begin{equation}
\label{eqn:const_b}
\alpha_{\rm 12} = \frac{\alpha_{\rm 11}\sigma_{\rm 11}+\alpha_{\rm 22}\sigma_{\rm 22}}{\sigma_{\rm 11}+\sigma_{\rm 22}},
\end{equation}
where $\sigma_{\rm 11}$ and $\sigma_{\rm 22}$ are the collisional cross-sections for 
direct collisions between two particles of type 1 and two particles of type 2, respectively.  The 
particles have identical masses, so the collisional cross-sections for these two collision scenarios 
are independent of mass and are set equal to the sum of their geometrical cross-sections, or 
$\sigma_{\rm 11} = 2\pi R_{\rm 1}^2$ (for example).  
Hence, from Equation~\ref{eqn:const_b}, the constant $\alpha_{\rm 12}$ is calculated as the 
mean of the constants $\alpha_{\rm 11}$ and $\alpha_{\rm 22}$, weighted by the corresponding 
geometrical cross-sections.  

Equations~\ref{eqn:Pcoll2} and~\ref{eqn:const_b} can be understood as follows.  During the 
evolution of a 
chaotic gravitational encounter, stars are continually undergoing close approaches with 
each other.  The probability of a close approach corresponding to a separation smaller than the sum 
of the particle radii occurring over the course of a crossing time is proportional to the number 
of interacting stars.  That is, over a given period of time, more 
close approaches occur if more particles are involved \citep{valtonen06}.  
However, if all particles are 
identical, then the probability of an event occurring (e.g. collision, ejection, etc.) 
in a crossing time is the same for all particles.  Hence, identical particles have the same 
probability of experiencing a close approach corresponding to a separation smaller than the 
sum of the particle radii (which we call a collision).  
If the particles are \textit{not} 
identical, then the volume of phase-space corresponding to a collision depends on the types of the 
colliding particles.  
The constants $\alpha_{\rm 11}$, $\alpha_{\rm 12}$ and $\alpha_{\rm 22}$ 
in Equation~\ref{eqn:const_b} correct 
for these different phase-space volumes for different collision scenarios, which we assume 
are (non-linearly) proportional to their respective collisional cross-sections.

The key point is that, knowing a priori the constants $\alpha_{\rm 11}$ and $\alpha_{\rm 22}$ (i.e. 
from numerical scattering experiments), the collision probabilities for the remaining $N-1$ different interaction 
scenarios\footnote{For a given $N$, there are 
$N+1$ total encounter scenarios, as is evident upon looking at Figure~\ref{fig:fig1}.  For the 2+3 case, this 
figure illustrates that there are six interaction scenarios in total (since five particles are involved 
in the interaction).  The coefficients for the two cases involving all identical particles are obtained directly 
from numerical scattering experiments.  Collision probabilities are then calculated 
for the four remaining interaction scenarios, with different numbers of stars of types 1 and 2, using these two coefficients.}
can be calculated directly without the need for any additional numerical scattering 
experiments.  As we will show in Section~\ref{results}, for a given total encounter energy and angular 
momentum, a relatively small set of numerical scattering experiments can be extrapolated from to 
directly obtain the coefficients $\alpha_{\rm 11}$ and $\alpha_{\rm 22}$ for \textit{any} particle 
radii.  Additionally, numerical scattering experiments need only be run for the 2+2 or 1+3 cases\footnote{In 
principle, the 1+2 case can even be used, but here the normalization in total energy and angular momentum is 
slightly more complicated (see \citealt{leigh12} for more details), since there is only one binary orbit going 
into the interaction.} since, 
using the binomial theorem, the corresponding coefficients for higher-$N$ interactions (2+3, 3+3, etc.) 
can be calculated directly.  This offers a strategy for using a minimal number of numerical scattering 
experiments to calculate collision probabilities for encounters involving any number of particles with 
any combination of particle radii.


\subsection{Numerical scattering experiments} \label{scatter}

In this section, we present the numerical scattering experiments used to test our strategy  
for calculating collision probabilities presented in the previous section.  

We calculate the outcomes of a series of binary-binary,
single-triple, binary-triple, and triple-triple
encounters using the \texttt{FEWBODY} numerical scattering code\footnote{for the source code, see
http://fewbody.sourceforge.net}.  The code
integrates the usual $N$-body equations in configuration- (i.e., position-)
space in order to advance the system forward in time.  For details
concerning the adaptive integration, classification techniques, etc. 
used by \texttt{FEWBODY}, we refer the reader to \citet{fregeau04}.

In Paper I, we adapted the \texttt{FEWBODY} code to handle encounters involving not only 
single and binary stars, but also triples.\footnote{Specifically we created additional 
subroutines to simulate 1+3 and 3+3 encounters; codes to simulate encounters between 
binaries and singles only, as well as a 2+3 encounter code, were previously available 
in the \texttt{FEWBODY} package.  The authors are happy to provide these additional subroutines 
to users of \texttt{FEWBODY} upon request.}  
We use the same criteria as \citet{fregeau04} to decide when a given
encounter is complete, defined as the point at which the
separately bound hierarchies that make up the system are no longer
interacting with each other or evolving internally.

To perform physical collisions between stars, \texttt{FEWBODY} uses the
``sticky-star'' approximation.  This treats stars as rigid spheres
with radii equal to their stellar radii.  A physical collision is
assumed to occur when the radii of the stars overlap.  When this
happens, the stars are merged assuming conservation of linear
momentum and no mass loss.  This does not account for tidal effects,
which could significantly increase the collision probability
\citep[e.g.][]{mcmillan86}, but are beyond the scope of this work.  For
this reason, we consider the collision probabilities presented in this
paper to be lower limits for the true values.

Previous scattering experiments have shown that the total energy and angular momentum 
are critical parameters in deciding the outcomes of 1+2 interactions 
\citep[e.g.][]{valtonen06}.  We confirmed this for higher-$N$ interactions (up to $N = 6$) 
in Paper I.  Therefore, we will fix the total energy and angular momentum 
when comparing between encounters involving different numbers of objects.
By fixing these quantities, we aim to remove the dependences of the encounter outcomes
on energy and angular momentum (when comparing between encounters involving different numbers 
of particles, or different combinations of particle types/radii), thereby normalizing the 
comparisons to reveal the dependences of the collision probability on other parameters, in 
particular the particle radii.  We consider one specific combination of the total energy 
and angular momentum (see below), for which both the total energy (range) and angular momentum
are always chosen to be the same to within a factor of $\sim 2$ for every type 
of encounter (see Paper I for more details).  

To this end, all orbits are circular, and have semi-major axes of either a$_{\rm 0} =$ 0.1 AU 
or a$_{\rm 1} =$ 1 AU.  Hence, for a 2+3 encounter for example, we adopt a$_{\rm 0} =$ 0.1 AU for the 
orbital separation of the inner orbit of the triple, and a$_{\rm 1} =$ 1 AU for both the outer 
orbit of the triple and the interloping binary.  For a 1+3 encounter, we similarly adopt 
a$_{\rm 0} =$ 0.1 AU and a$_{\rm 1} =$ 1 AU for the inner and outer orbits, respectively, 
of the triple, while for a 2+2 encounter we adopt a$_{\rm 0} =$ 0.1 AU and a$_{\rm 1} =$ 1 AU for 
the two binaries.  These initial semi-major axes are summarized in Table~\ref{table:initial-conditions}.  
A semi-colon is used to separate different objects, whereas a comma is used to separate the orbits within 
triples.  Parantheses are used to enclose the semi-major axes of triples, with the smaller of the two 
semi-major axes always corresponding to the inner binary.  
Note that we have checked that our 
results are approximately insensitive to our choice of semi-major axes, for orbital separation ratios 
$a_{\rm 0}$/$a_{\rm 1} \lesssim$ 100.  For a given relative velocity at infinity and impact parameter, 
the choice of semi-major axes sets the encounter energy and angular momentum.  Thus, while the exact 
collision probability does depend on the choice of semi-major axes, our method does not.  
This choice of semi-major axes, along with our chosen impact parameter ($b =$ 0; see below), serves to
minimize the computational expense of our scattering experiments, by
minimizing the total angular momentum.  Regardless, these semi-major axes are relatively characteristic of 
the multiple star systems we expect to be present in globular clusters.  At least in the cluster core, 
only relatively compact multiples should survive for several Gyr in such dense stellar environments.  
This is especially true for triples (and higher-order multiples), given that 
they are only stable against dissociation if the ratio between their inner and outer orbital 
separations is sufficiently small \citep[e.g.][]{mardling01}.

\begin{table*}
\caption{Initial semi-major axes of all binaries and triples}
\begin{tabular}{|c|c|}
\hline
Encounter Type   &      Semi-major axes      \\
                 &         (in AU)           \\
\hline
2+2              &         0.1; 1.0           \\
1+3              &        (0.1, 1.0)          \\
2+3              &     1.0; (0.1, 1.0)       \\
3+3              &  (0.1, 1.0); (0.1, 1.0)   \\
\hline
\end{tabular}
\label{table:initial-conditions}
\end{table*}

For each Run, we populate a grid of scattering experiments varying only the
relative velocity at infinity $v_{\rm inf}$.  Specifically, we select values for the relative velocity 
at infinity from 0 to 1.1 in equally-spaced intervals of 0.004,
in units of the critical velocity $v_{\rm crit}$ (defined as the relative velocity that gives a
total energy of zero for the encounter).  Hence, our chosen range of relative velocities 
at infinity includes the case of zero total energy.  In the case v$_{\rm inf} >$ v$_{\rm crit}$, the encounter is 
prompt and at least one particle is almost immediately lost from the system.  For v$_{\rm inf} <$ v$_{\rm crit}$, 
all particles may be temporarily bound and participate in a prolonged resonant encounter.  
We fix the impact parameter 
at $b = 0$, independent of $v_{\rm inf}$.  For our assumed circular orbits, this minimizes the 
total angular momentum (and hence reduces the total computational cost of our simulations) while 
preserving our normalization that the total energy and angular 
momentum must be roughly the same independent of the number of interacting particles.  
Finally, at each point in this grid, we perform multiple scattering experiments
randomizing all angles in the encounter, including the angles at impact between the 
orbital planes of any binaries and triples involved in the encounters.  In total, 
we perform 24750 simulations for every Run (with the exception of Figure~\ref{fig:fig4}, 
for which we are limited to using only 4500 simulations per data point). 
We chose this number of simulations to balance between 
statistical significance and computational expense (i.e. simulation run-time).  
Of course, running  more simulations would reduce the uncertainties on the calculated 
collision probabilities and yield more precise results, which may be desirable for certain applications.

The collision probability is calculated from the output of our numerical
scattering experiments for a given encounter type and a given Run as:
\begin{equation}
\label{eqn:prob_coll_scat}
P_{\rm coll} = \frac{N_{\rm coll}}{N_{\rm tot}},
\end{equation}
where $N_{\rm coll}$ is the total number of encounters that result in a direct
physical collision, and $N_{\rm tot}$ is the total number of encounters performed 
(throughout this paper, $N_{\rm tot} = 24750$,
except for Figure~\ref{fig:fig4}, as stated above).  The uncertainty for the collision 
probability is calculated using Poisson statistics according to:
\begin{equation}
\label{eqn:prob_coll_err}
{\Delta}P_{\rm coll} = \Big( \Big( \frac{\sqrt{N_{\rm coll}}}{N_{\rm tot}} \Big)^2 + {\Delta}P_{\rm coll,int}^2 \Big)^{1/2},
\end{equation}
where ${\Delta}P_{\rm coll,int}$ is an additional term to account for any intrinsic dispersion in 
our calculated collision probabilities (see Section~\ref{results} for our assumed values of 
${\Delta}P_{\rm coll,int}$).  This additional source of dispersion comes from 
a combination of
numerical inaccuracies, our normalization in energy and angular momentum, and subtle inadequacies 
of our model in capturing all of the relevant underlying physics (e.g. the issue of saturation 
which we discuss in more detail in Section~\ref{results}).\footnote{Formally, the intrinsic 
dispersion term ${\Delta}P_{\rm coll,int}$ should be found by forcing the reduced chi-square 
of a fit to be unity.  However, in our case, we have only three data points (i.e. $N =$ 4, 5 and 6) 
to work with in calculating the intrinsic dispersion term.  Thus, necessarily, our estimates 
for ${\Delta}P_{\rm coll,int}$ are only approximate.}  
Technically, this uncertainty should include scattering experiments that result in
unresolved outcomes \citep{hut83b}.  However, we find that the number of unresolved
outcomes is sufficiently small that it does not significantly contribute to
${\Delta}P_{\rm coll}$, and we do not include it in its calculation.  

\section{Results} \label{results}

In this section, we first present some example applications of our method.  Using 
these, we go on to describe how to calculate collision probabilities for any number 
of particles and any combination of particle radii, as well as for a limited 
range of particle masses.

%
\subsection{Different particle radii} \label{radii}

\begin{figure}
\begin{center}
\includegraphics[width=\columnwidth]{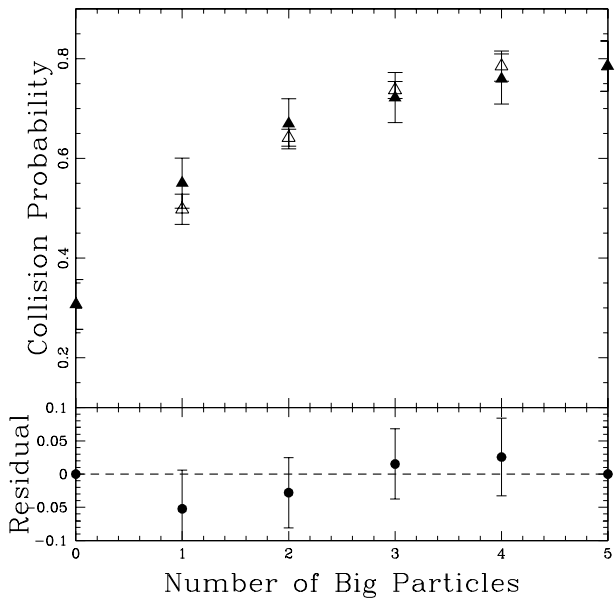}
\end{center}
\caption[Collisional probabilities for binary-triple encounters as a function of the number of large particles.]{The 
probability of a collision occurring during a binary-triple encounter, shown in the top panel.  Two particle radii are 
considered, namely 
$R_{\rm 1} =$ 0.1 R$_{\odot}$ and $R_{\rm 2} =$ 5 R$_{\odot}$.  The number of large particles is shown on the $x$-axis, 
and the collision probability is shown on the $y$-axis.  All particle masses are identical, and the range of 
total encounter energies and angular momenta over which we integrate are identical for all cases (i.e. combinations 
of small and large particles).  We adopt $a_{\rm 0} =$ 0.1 AU for the orbital separation of the inner orbit of the 
triple, and $a_{\rm 1} =$ 1 AU for both the outer orbit of the triple and the interloping binary.  All orbits have 
eccentricities of zero, initially.  The open triangles 
show our predictions for the collision probabilities, calculated 
from Equations~\ref{eqn:Pcoll2} and~\ref{eqn:const_b} using the results of our numerical scattering experiments for 
the $N_{\rm 1} = 0$ and $N_{\rm 2} = 5$ cases (i.e. all small and all large particles).  The filled triangles show the same collision 
probabilities, but calculated directly from the results of the corresponding numerical scattering experiments.  Error 
bars are calculated using Poission statistics (filled triangles) and simple error propagation (open triangles), 
beginning with Equation~\ref{eqn:prob_coll_err} and assuming ${\Delta}P_{\rm coll,int} =$ 0.05.  The bottom
panel shows the residuals between the collision probabilities calculated directly from our numerical
scattering experiments (filled triangles) and our model predictions (open triangles).  The dashed line is shown
for reference, and represents perfect agreement between our model and the data.
\label{fig:fig1}}
\end{figure}

First, consider the case where the particles all have the same mass, but two different 
particle radii are possible.  We assume $R_{\rm 1} =$ 0.1 R$_{\odot}$ and $R_{\rm 2} =$ 5 R$_{\odot}$, 
and $m_{\rm 1} = m_{\rm 2} =$ 1 M$_{\odot}$.  Figure~\ref{fig:fig1} shows the results of 
a series of numerical scattering simulations of binary-triple encounters (i.e. $N = 5$).  
Here, we vary the number of large particles from 0 to 5, as shown on the $x$-axis.  The 
constants $\sigma_{\rm 11}$ and 
$\sigma_{\rm 22}$ in Equation~\ref{eqn:Pcoll2} are calculated from our numerical scattering 
experiments using the $N_{\rm 1} = 5$ and $N_{\rm 2} = 5$ cases, where all particles are either of type 1 
($\alpha_{\rm 11}$) or type 2 ($\alpha_{\rm 22}$).  The constant $\alpha_{\rm 12}$ is then 
calculated from Equation~\ref{eqn:const_b}.  The open points show the results of these 
calculations, and are compared directly to our numerical scattering experiments via the 
filled points.  Error bars are calculated using Equation~\ref{eqn:prob_coll_err} with 
${\Delta}P_{\rm coll,int} = 0.05$, which is chosen to ensure that the residuals calculated in 
Figure~\ref{fig:fig1} are all within one standard deviation of zero.  Our calculated predictions agree 
very well with the results of our numerical scattering experiments, and are within one standard 
deviation of each other in all cases.  
The agreement is slightly worse 
for the $N_{\rm 2} = 1$ and N$_{\rm 2} = 2$ cases, however, for which our calculated collision probabilities 
deviate from our numerical scattering experiments 
by about one standard deviation (see the bottom panel in Figure~\ref{fig:fig1}).  The reason for this 
poor agreement is not understood.  This may be highlighting that we are missing 
some higher-order term(s)
in 
Equation~\ref{eqn:Pcoll2}, possibly related to our calculation of the coefficient 
$\sigma_{\rm 12}$.  
Regardless, the calculated collision 
probabilities (via Equation~\ref{eqn:Pcoll2}) always agree with the results of 
our numerical scattering experiments to within a couple percent or better. 

\begin{figure}
\begin{center}
\includegraphics[width=\columnwidth]{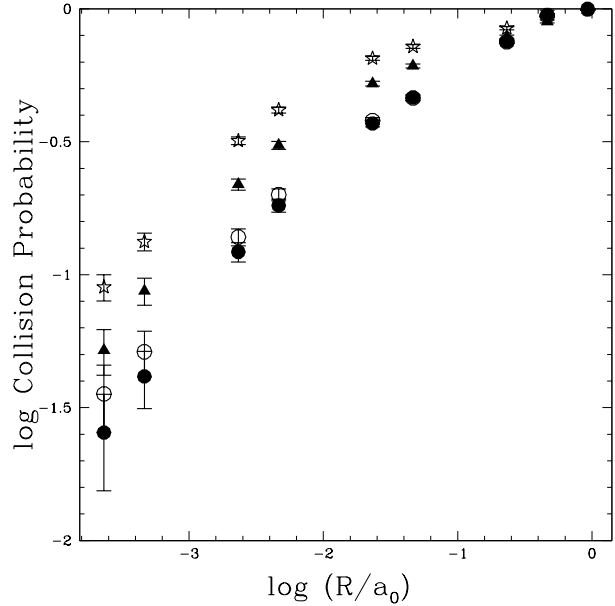}
\end{center}
\caption[Collision probabilities for 2+2, 1+3, 2+3 and 3+3 encounters as a function of the ratio $R$/$a_{\rm 0}$]{The 
logarithm of the probability of a collision occurring during encounters involving all identical particles is shown 
as a function of the logarithm of the ratio $R$/$a_{\rm 0}$, where $R$ is the particle radius in AU and $a_{\rm 0} =$ 0.1 AU 
is the initial semi-major axis of the most compact orbit going into the encounter.  
Hence, we adopt $a_{\rm 0} = 0.1$ AU for the orbital separations of the 
inner orbits of any triples, and $a_{\rm 1} = 1$ AU for the outer orbits of any triples.  For the 2+3 case, the initial 
semi-major axis of the binary is $a_{\rm 1} = 1$ AU, whereas for the 2+2 case the binaries have different initial 
semi-major axes of $a_{\rm 0} = 0.1$ AU and $a_{\rm 1} = 1$ AU.  Thus, $a_{\rm 0}$ is a constant for all encounters, 
independent of the total number of particles or the particle radii.  The filled circles, open circles, filled triangles 
and open stars correspond to 2+2, 1+3, 2+3 and 3+3 encounters, respectively.  Error bars are calculated using 
Equation~\ref{eqn:prob_coll_err} with ${\Delta}P_{\rm coll,int} =$ 0.01. 
\label{fig:fig2}}
\end{figure}

In Figure~\ref{fig:fig2} we show the results of a series of 2+2 (filled circles), 1+3 (open circles), 
2+3 (filled triangles) and 3+3 (open stars) encounters, this time involving all identical particles.  The 
encounter parameters are the same as in Figure~\ref{fig:fig1} 
and Table~\ref{table:initial-conditions}, with the sole exception that the particle 
radii are all identical.  For each encounter type, we show the collision 
probability as a function of the ratio $R$/$a_{\rm 0}$, where $R$ is the particle radius 
and $a_{\rm 0}$ is the initial semi-major axis of the most compact orbit going into the 
encounter (i.e. $a_{\rm 0} =$ 0.1 AU).  Hence, $R$/$a_{\rm 0} = 1$ corresponds to a contact 
configuration for the most compact orbit initially. 

The key point is that an analogous version of Figure~\ref{fig:fig1} can be made for any combination of 
particle radii (i.e. any $R_{\rm 1}$ and $R_{\rm 2}$).  In principle, this can be done by extracting the coefficients 
$\sigma_{\rm 11}$ and $\sigma_{\rm 22}$ directly from Figure~\ref{fig:fig2}, 
and fitting a curve to these data to derive the collision probability at any value of $R$ (and $N$).
The temptation is, for every $N$, to fit the data with a function of the form:
\begin{equation}
\label{eqn:func}
P_{\rm coll} = {\delta}R^{\gamma}{N \choose 2}
\end{equation}
However, as is clear from Figure~\ref{fig:fig2}, a plot of $R$ versus $P_{\rm coll}$ is clearly 
sub-linear.  Even in log-log space, significant curvature is apparent.  
This is not unexpected, since in the limits $R \rightarrow 0$ and $R \rightarrow a_{\rm 0}$ (where $a_{\rm 0}$ 
is the semi-major axis of the most compact orbit going into the encounter) 
we expect, respectively, $P_{\rm coll} \rightarrow 0$ and $P_{\rm coll} \rightarrow 1$, independent 
of the number of interacting particles $N$.  
Thus, the coefficients $\delta$ and 
$\gamma$ in Equation~\ref{eqn:func} are not constants, but are instead functions of the particle radius $R$.  
(In practice, one could fit a function similar to Equation~\ref{eqn:func} to a subset of data around the desired value of 
$R/a_0$, in order to derive the collision probability at a point not plotted in Figure~\ref{fig:fig2}).

\begin{figure}
\begin{center}
\includegraphics[width=\columnwidth]{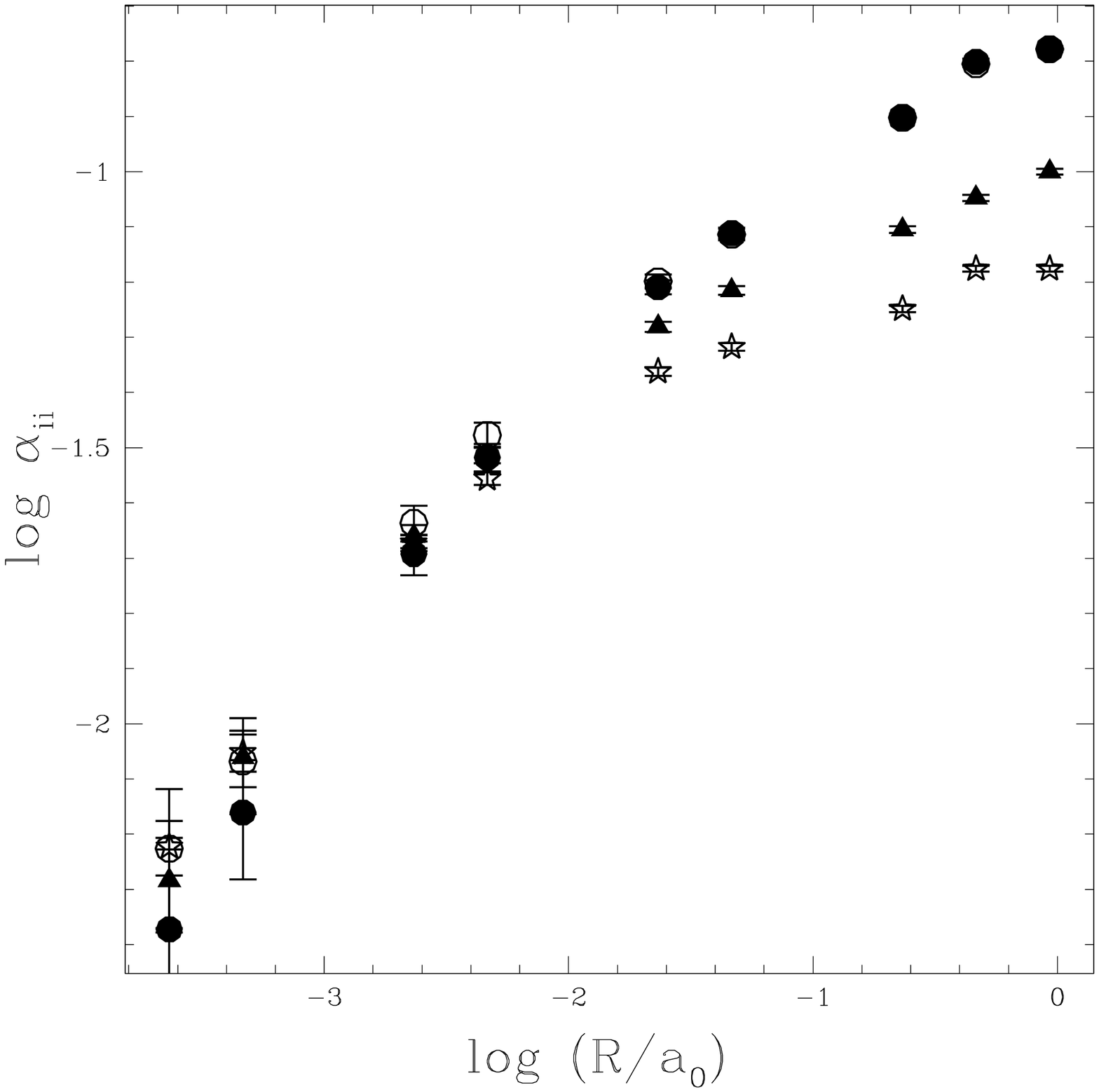}
\end{center}
\caption[The coefficients $\alpha_{\rm ii}$ for encounters involving all identical particles as a function 
of the particle radius]{The logarithm of the coefficient $\alpha_{\rm ii}$ given in Table~\ref{table:one}, calculated as 
$P_{\rm coll}$/${N \choose 2}$, as a function of the logarithm of the ratio $R$/$a_{\rm 0}$, where $R$ is the 
particle radius in AU and $a_{\rm 0} =$ 0.1 AU is the initial semi-major axis of the most compact orbit going into the 
encounter, for encounters involving all identical particles.  We assume the same encounter parameters and 
symbols as in Figure~\ref{fig:fig2}.  Error bars are calculated as in Figures~\ref{fig:fig1} and~\ref{fig:fig2}.
\label{fig:fig3}}
\end{figure}

Since here, by construction, the total encounter energy and angular 
momentum are roughly the same independent of $N$, then, via the binomial theorem, at a given value of the 
particle radius $R$, 
$\delta$ and $\gamma$ should be the same for all $N$.  That is, since the 
particles are all identical and the total encounter energy and angular momentum are fixed, the collision 
probability should vary only with the number of interacting particles $N$.
To illustrate this point, we calculate the quantity $P_{\rm coll}$/${N \choose 2} = \alpha_{\rm ii}$ for encounters 
involving all identical particles, for 
all values for the particles radius, $R$, shown in Figure~\ref{fig:fig2}, 
with the expectation that the ratio $P_{\rm coll}$/${N \choose 2}$ will be the same for all $N$ (to within 
the uncertainties).  The results are shown in Table~\ref{table:one} and Figure~\ref{fig:fig3}.  
Uncertainties are calculated using Equation~\ref{eqn:prob_coll_err} with ${\Delta}P_{\rm coll,int} =$ 0.01, 
which ensures that our simulated collision probabilities for the 2+2 and 1+3 cases are always within one 
standard deviation of each other.

The quantity $P_{\rm coll}$/${N \choose 2}$ is approximately the same (e.g., within typically less than one standard deviation) 
for all $N$ at a given particle radius $R$, until reaching a ``saturation'' limit at $\log (R/a_0) \gtrsim -1.5$.  This suggests that, 
at least for a given particle radius $R$, the collision probabilities for 2+3 and 
3+3 encounters can be calculated directly from the collision probability for the 2+2 (or 1+3) case, without having 
to run any 2+3 and 3+3 numerical scattering experiments.  This offers the 
potential to calculate collision probabilities for high-$N$ interactions with minimal computational expense, which 
increases steeply with increasing $N$ (as we also point out in Paper I).  Equivalently, for fixed computational 
expense a larger number of simulations can be performed for lower $N$ interactions, which subsequently increases 
the statistical significance and accuracy of the results.  We caution that the simulated data begin to deviate from 
the trend that $P_{\rm coll}$/${N \choose 2}$ is roughly independent of $N$ near the saturation limit 
$P_{\rm coll} \rightarrow 1$.  As shown in Figure~\ref{fig:fig3}, this begins around $\log (R/a_0) \sim -1.5$,
and should be treated with caution.  More work 
is needed to verify that this exact saturation limit is ubiquitous for all total encounter energies and angular 
momenta.  In general, however, adopting larger initial semi-major axes $a_{\rm 0}$ pushes the saturation limit to 
larger values of the particle radius $R$. 

\begin{table*}
\caption{The quantity $P_{\rm coll}$/${N \choose 2}$ as a function of the particle radius R}
\begin{tabular}{|c|c|c|c|c|}
\hline
Particle Radius   &  \multicolumn{4}{|c|}{$P_{\rm coll}$/${N \choose 2}$}         \\
 (in R$_{\odot}$) &         (2+2; $N$=4)         &             (1+3; $N$=4)       &          (2+3; $N$=5)       &          (3+3; $N$=6)        \\
\hline
 0.005          &  0.0042 $\pm$ 0.0017  &  0.0059 $\pm$ 0.0017  &  0.0052 $\pm$ 0.0010  &  0.0060 $\pm$ 0.0007   \\
 0.01           &  0.0069 $\pm$ 0.0017  &  0.0085 $\pm$ 0.0017  &  0.0087 $\pm$ 0.0010  &  0.0089 $\pm$ 0.0007   \\
 0.05           &  0.0203 $\pm$ 0.0017  &  0.0231 $\pm$ 0.0017  &  0.0219 $\pm$ 0.0010  &  0.0213 $\pm$ 0.0007   \\
 0.1            &  0.0304 $\pm$ 0.0017  &  0.0333 $\pm$ 0.0017  &  0.0307 $\pm$ 0.0011  &  0.0278 $\pm$ 0.0007   \\
 0.5            &  0.0618 $\pm$ 0.0018  &  0.0633 $\pm$ 0.0018  &  0.0524 $\pm$ 0.0011  &  0.0434 $\pm$ 0.0007   \\
 1.0            &  0.0773 $\pm$ 0.0018  &  0.0770 $\pm$ 0.0018  &  0.0610 $\pm$ 0.0011  &  0.0481 $\pm$ 0.0008   \\
 5.0            &  0.1255 $\pm$ 0.0019  &  0.1252 $\pm$ 0.0019  &  0.0785 $\pm$ 0.0011  &  0.0564 $\pm$ 0.0008   \\
10.0            &  0.1581 $\pm$ 0.0020  &  0.1567 $\pm$ 0.0020  &  0.0896 $\pm$ 0.0012  &  0.0667 $\pm$ 0.0008   \\
20.0            &  0.1667 $\pm$ 0.0020  &  0.1667 $\pm$ 0.0020  &  0.1000 $\pm$ 0.0012  &  0.0667 $\pm$ 0.0008   \\
\hline
\end{tabular}
\label{table:one}
\end{table*}


\subsection{Different particle masses} \label{masses}

\begin{figure}
\begin{center}
\includegraphics[width=\columnwidth]{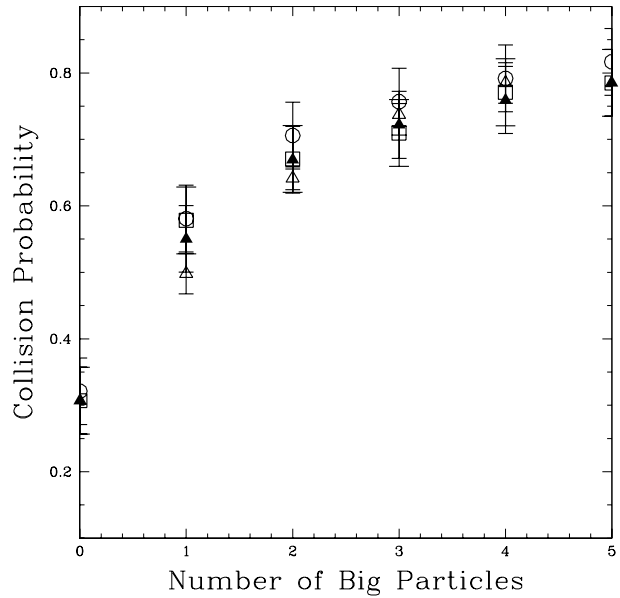}
\end{center}
\caption[Collisional probabilities for binary-triple encounters involving different particle masses as a function 
of the number of large particles.]{The 
probability of a collision occurring during a binary-triple encounter.  
We assume the same encounter parameters and particle radii as in Figure~\ref{fig:fig1}.  This time, however, the particle masses 
can be different.  The open and filled triangles show the same collision probabilities as in Figure~\ref{fig:fig1}, and 
assume all equal mass particles (i.e. 1 M$_{\odot}$).  That is, the open triangles are calculated
from Equations~\ref{eqn:Pcoll2} and~\ref{eqn:const_b} using the results of our numerical scattering experiments for
the $N_{\rm 1} = 0$ and $N_{\rm 2} = 5$ cases only, and the filled triangles show the same collision
probabilities but calculated directly from the results of the corresponding numerical scattering experiments.
The open circles show the same encounter parameters as in Figure~\ref{fig:fig1} and are calculated directly 
from our numerical scattering experiments, but assuming particle masses 
of 0.7, 0.8, 0.9, 1.0 and 1.0 M$_{\odot}$ for every simulation.  The open squares again show the same 
encounter parameters and are calculated directly from our numerical scattering experiments, but adopting 
radii and masses of $R_{\rm 1} =$ 0.1 R$_{\odot}$ and $m_{\rm 1} =$ 0.5 M$_{\odot}$ for particles of type 1, 
while particles of type 2 have radii $R_{\rm 2} =$ 5 R$_{\odot}$ and 
masses $m_{\rm 2} =$ 1 M$_{\odot}$.  Error bars are calculated as in Figure~\ref{fig:fig1} (i.e. using 
Equation~\ref{eqn:prob_coll_err} with ${\Delta}P_{\rm coll,int} =$ 0.05).
\label{fig:fig4}}
\end{figure}

Next, we consider encounters involving particles having different radii \textit{and} different masses.  The 
key point we wish to illustrate in this section is that, provided the range of energies and angular 
momenta are fixed and independent of $N$, Equation~\ref{eqn:Pcoll1} is approximately 
valid and describes the collision probabilities to within a few percent, \textit{independent of 
the distribution of particle masses, provided they are all within a factor $\lesssim 2$.}  
Figure~\ref{fig:fig4} shows two such examples.  The first is shown by the open triangles, 
for which the particle masses are fixed and independent of our choices for the particle 
radii.  We set the particle masses to 0.7, 0.8, 0.9, 1.0 and 1.0 M$_{\odot}$, with particle 
radii of either $R_{\rm 1} =$ 0.1 R$_{\odot}$ or $R_{\rm 2} =$ 5 R$_{\odot}$ assigned randomly (i.e. 
independent of particle mass).  The 
initial configuration (i.e. which objects form the inner binary of the triple, etc.) is also 
chosen at random.  
The second 
example is shown by the open squares.  Here, we vary the particle masses along with the 
particle radii, such that particles of type 1 have R$_{\rm 1} =$ 0.1 R$_{\odot}$ and 
m$_{\rm 1} =$ 0.5 M$_{\odot}$, while particles of type 2 have R$_{\rm 2} =$ 5 R$_{\odot}$ and
m$_{\rm 2} =$ 1 M$_{\odot}$.  Again, the initial configuration is then chosen at random.  
Figure~\ref{fig:fig4} shows that in both of these examples, 
Equation~\ref{eqn:Pcoll1} accurately describes the collision probabilities to 
within a few percent.  We have checked that this is also the case for mass ratios $\sim 0.05$, 
however a more thorough exploration of the relevant parameter space is required to accurately 
determine to what mass ratios this approximation holds.  Our results support the conclusion that 
it is valid for mass ratios $\gtrsim 0.5$, and possibly much lower.

\subsection{Putting it all together} \label{put}

In this section, we combine the results of the previous two sections to construct a simple 
procedure for calculating collision probabilities in realistic star clusters, for any 
combination of particle radii and masses.  Thus far, we have focussed on encounters involving 
only two different types of particles.  The method presented below, however, is more general.


To summarize, our method is as follows:
\begin{itemize}
\item Decide upon the encounter type and, in particular, the total number 
of particles involved in the interaction.  As a guide, the observed fractions 
of binary and multiple star systems can be used to calculate analytic estimates 
for the rates of the different encounter types 
\citep[e.g. 2+2, 2+3, etc., see ][]{leonard89,leigh11}.  
\item 
Determine
the initial total encounter energy and angular momentum
from the initial orbital eccentricities and semi-major axes, along with the
initial relative velocity at infinity and impact parameter.  Note that the method
is valid 
both for a particular combination of energy and angular momentum,
as well as consistent ranges in these quantities.  
Depending on how much information is available, one might estimate the relative velocities at infinity and the impact parameter in a star cluster
from the observed surface brightness and velocity dispersion profiles \citep[e.g.][]{leonard89,bahramian13}.
Likewise, binary orbital parameters and masses can be drawn from the appropriate observed (or assumed) distribution functions.
\item Set the particle types, in particular the particle radii $R_{\rm i}$, along with 
the particle masses.  Note that, as shown in Section~\ref{masses}, our method is valid 
to within a few percent for any combination of particle masses (to within a factor 
of a few).\footnote{In a real star cluster, the distribution of stellar masses and radii is 
continuous.  Hence, some binning by stellar mass (which is tightly correlated with stellar 
radius on the main-sequence) is desirable, wherever this assumption is reasonable.  For example, 
binning is a reasonable assumption for interactions involving white dwarfs, bright red 
giants and/or main-sequence stars close to the turn-off.  Here, the objects all have 
very similar radii and masses \textit{within} their individual stellar evolution-based 
classifications, but very different radii \textit{between} these classifications.}
\item Given the particle radii, obtain the coefficients $\alpha_{\rm ii}$, 
$\alpha_{\rm jj}$, etc. directly from numerical scattering experiments, 
using the parameters determined above.  
With these, calculate the coefficient(s) $\alpha_{\rm ij}$ using the generalized 
equation:
\begin{equation}
\label{eqn:alphaij}
\alpha_{\rm ij} = \frac{\alpha_{\rm ii}\sigma_{\rm ii}+\alpha_{\rm jj}\sigma_{\rm jj}}{\sigma_{\rm ii}+\sigma_{\rm jj}},
\end{equation}
where $\sigma_{\rm ii} = 2{\pi}R_{\rm i}^2$.  As shown in Figure~\ref{fig:fig2}, nearly all possible values 
of the coefficients 
$\alpha_{\rm ii}$, $\alpha_{\rm jj}$, etc.\ can be covered using a limited number of numerical scattering 
experiments by extrapolating between data points.  Additionally, numerical scattering experiments need only 
be run for the 2+2 or 1+3 cases since, using the binomial theorem, the corresponding coefficients for higher-$N$ 
interactions (2+3, 3+3, etc.) can be calculated directly (e.g., see end of Section~\ref{radii}).  
This significantly reduces the total computational 
expense for higher-$N$ interactions, since the integration times for the simulations increase dramatically 
with increasing particle number.
\item Calculate the desired collision probabilities using the generalized 
formula:
\begin{equation}
\label{eqn:Pcollgen}
P_{\rm coll} = \sum_{\rm i=1}^{\rm N_{\rm i}}\alpha_{\rm ii}{N_{\rm i} \choose 2} + \frac{1}{2}\sum_{\rm i=1}^{\rm N_{\rm i}}\sum_{\rm j=1,i{\ne}j}^{\rm N_{\rm j}}\alpha_{\rm ij}{N_{\rm i} \choose 1}{N_{\rm j} \choose 1}
\end{equation}
\end{itemize}

As a proof of concept of the above generalized methodolgy, consider 2+3 encounters involving three 
different particle types characterized by their different radii, or $R_{\rm 1} =$ 0.05 R$_{\odot}$,
$R_{\rm 2} =$ 0.5 R$_{\odot}$ and $R_{\rm 3} =$ 5.0 R$_{\odot}$.  All particles have
the same mass of 1 M$_{\odot}$, and the semi-major axes and eccentricities are chosen to be
the same as used in Figure~\ref{fig:fig2} and Table~\ref{table:one}, as is the range of relative 
velocities at infinity (including a small fraction of encounters with positive total energies).  
As before, the impact parameter is set equal to zero for all simulations.  
We consider five different combinations of particle types, and run 24750 simulations for each 
such combination.  For comparison, we calculate the corresponding collision probabilities (along with 
their uncertainties $\sigma P_{\rm coll}$) using Equations~\ref{eqn:Pcollgen} 
and~\ref{eqn:alphaij} as well as the coefficients $\alpha_{\rm ii}$ and $\alpha_{\rm jj}$ obtained 
from Table~\ref{table:one}.  Based 
on Figure~\ref{fig:fig3}, $R_{\rm 3} =$ 5 R$_{\odot}$ is beyond the saturation limit for our
model.  This will introduce some small amount of additional scatter into our calculations for 
the collision probabilities.  To account for this additional source of uncertainty, we calculated
the mean value of $\alpha_{\rm ii}$ at $R =$ 5 R$_{\odot}$ in Figure~\ref{fig:fig3} for all values 
of the particle number $N$ (i.e. $N =$ 4, 5 and 6), and added the
resulting value to its uncertainty (in quadrature) using Equation~\ref{eqn:prob_coll_err} with 
${\Delta}P_{\rm coll,int} =$ 0.03.  
This serves to illustrate the importance of accurately estimating
the coefficients $\alpha_{\rm ii}$, $\alpha_{\rm jj}$, etc. when calculating collision probabilities.

The results of this example are shown in Table~\ref{table:two}.  The agreement between our simulated 
and calculated collision probabilities is quite good in all cases, which generally agree to within a few 
percent.  

\begin{table*}
\caption{The collision probability during 2+3 encounters for various combinations of particle types 1, 2 and 3}
\begin{tabular}{|c|c|c|c|c|}
\hline
\multicolumn{3}{|c|}{Number of Each Type of Particle}   &  \multicolumn{2}{|c|}{Collision Probability}         \\
\hline
$N_{\rm 1}$ & $N_{\rm 2}$ & $N_{\rm 3}$ &  Simulated  &  Calculated  \\
$R_{\rm 1} =$ 0.05 R$_{\odot}$  &  $R_{\rm 2} =$ 0.5 R$_{\odot}$ & $R_{\rm 3} =$ 5.0 R$_{\odot}$ & $P_{\rm coll}$ $\pm$ $\sigma P_{\rm coll}$  &  $P_{\rm coll}$ $\pm$ $\sigma P_{\rm coll}$  \\
\hline
4  &  1  &  0  &  0.3324 $\pm$ 0.0042  &  0.3398 $\pm$ 0.0074 \\
4  &  0  &  1  &  0.5264 $\pm$ 0.0057  &  0.4454 $\pm$ 0.1397 \\
3  &  2  &  0  &  0.4084 $\pm$ 0.0048  &  0.4307 $\pm$ 0.0073 \\
3  &  1  &  1  &  0.5670 $\pm$ 0.0060  &  0.5357 $\pm$ 0.1103 \\
1  &  1  &  3  &  0.7221 $\pm$ 0.0071  &  0.7578 $\pm$ 0.1807 \\
\hline
\end{tabular}
\label{table:two}
\end{table*}

\section{Summary and discussion} \label{summary}


In this paper, we present a formalism for calculating collision probabilities during chaotic gravitational 
encounters with finite-sized particles.  The interactions involve any number of particles and any combination of 
particle radii, using a minimal number of numerical scattering experiments.  Our method relies on 
the binomial theorem, and can also be extended to encounters involving particles of different masses (see 
below).  

Our model is applicable for a constant value or range in the total encounter energy and 
angular momentum.  Hence, our formalism for the collision probability can be extended to include 
any combination of total encounter energy and angular momentum.  To do this, plots analogous to 
Figure~\ref{fig:fig2} are generated for different total encounter energies at fixed 
angular momentum (and vice versa).   From here, 
the dependence of the fitting parameters (see 
Equation~\ref{eqn:func}) on the total encounter energy and angular momentum is derived.  
In general, the collision probability decreases with increasing total encounter energy (i.e. becomes 
less negative and thus closer to zero), and with increasing total angular momentum.  
Depending on the difficulties associated with any curvature in log-log space (which prohibits 
fitting a constant power-law to the data), an analytic formula for the collision probability can be 
derived by running additional numerical scattering experiments to cover the full range in encounter 
energies and angular momenta.  The resultant formula would thus cover the entire range of the 
relevant parameter space (total encounter energy and 
angular momentum, distribution of particle masses and radii, particle number, etc.,
in, for example, open, globular or nuclear star cluster clusters).  Alternatively, our method can 
be used to create a library of collision 
probabilities, again covering the entire range of the relevant parameter space 
while minimizing the total number of numerical scattering experiments that need to be performed 
(by extrapolating between data points, as in Figure~\ref{fig:fig2}).  This will be the focus 
of future work.

Our formalism is valid as well for different particle masses, provided they are all within a 
factor of $\sim$ a few.  This is because changing the particle masses also changes the total 
encounter energy and angular momentum.  Thus, by fixing the total energy and angular 
momentum for all encounters, the dependence of the collision probability is removed, at 
least to first-order.  There are second-order effects related to the lifetime of the system, 
causing slightly higher collision probabilities for encounters with large particle mass 
ratios.  However, the deviation is by at most a few percent from the model presented in this 
paper, which assumes all equal mass particles.
Our method is ideal for application in old open and globular clusters, since it is applicable 
over a wide range of particle radii but a relatively narrow range of particle masses.  Suitable 
interactions occur in GCs when single, binary and triple stars undergo encounters, since 
the objects involved include main-sequence stars (close to the turn-off), red or asymptotic giant 
branch stars, white dwarfs and neutron stars.  All of these objects have masses that 
are typically the same to within a factor $\lesssim$ 2, but radii differing by several 
orders of magnitude.  Thus, the geometric cross-section dominates the total collisional 
cross-section for these objects, as opposed to the gravitationally-focused cross-section.  
For example, consider an average encounter in a typical GC, involving mostly 
white dwarfs.  If even a single main-sequence star is included in the interaction, then 
the collision probability can increase by nearly an order of magnitude (this is the case for, e.g. 
the encounter 
parameters assumed in Figure~\ref{fig:fig2}, for which the collision probability begins to saturate 
at the particle radius corresponding to that of a typical main-sequence star in an old globular 
cluster).\footnote{The coefficients $\alpha_{\rm 11}$ and $\alpha_{\rm 22}$ for 
main-sequence stars and white dwarfs, respectively, are obtained from Table~\ref{table:one}, and the 
coefficient $\alpha_{\rm 12}$ is then calculated using Equation~\ref{eqn:const_b}.  The collision 
probability is subsequently calculated using Equation~\ref{eqn:Pcoll2}.}   
Consider a 2+2 interaction with the same encounter parameters as adopted in Figure~\ref{fig:fig2} and 
Table~\ref{table:one}.  If the interaction involves only white dwarfs, then P$_{\rm coll} \sim$ 0.04.  
If one of the white dwarfs is replaced by a single (near turn-off) main-sequence star, 
then P$_{\rm coll} \sim$ 0.25.
Conversely, consider the same encounter but involving only main-sequence stars with masses 
close to the turn-off mass.  The collision probability for this encounter is 
P$_{\rm coll} \sim$ 0.46, roughly an order of magnitude larger than calculated for the same 
encounter but involving 
only white dwarfs.  Here, if a single main-sequence star is replaced by a white dwarf, 
then the collision probability changes to P$_{\rm coll} \sim$ 0.31.

%

\section*{Acknowledgments}

We would like to thank Christian Knigge for useful discussions and suggestions.  
N.~W.~C.~L. is grateful for the generous support of an NSERC Postdoctoral Fellowship.  
A.~M.~G. is funded by a National Science Foundation Astronomy and Astrophysics Postdoctoral Fellowship under Award No.\ AST-1302765.


\bsp

\label{lastpage}

\end{document}